# Integrated routing for a vehicle-robot pickup and delivery system with time constraints

Yongjian Li[†], Yan Chen[†], Gaicong Guo[†], Huiwen Wu[†] and Zhao Yuan[*]
[†]*Department of Decision Sciences*
*Macau University of Science and Technology, Macau, China*
[*]*Electrical Power Systems Laboratory (EPS-Lab)*
*University of Iceland, Reykjavik, Iceland*
Email: zhaoyuan@hi.is, zhaoyuan.epslab@gmail.com

*Abstract*—This paper considers an unmanned vehicle-robot pickup and delivery system, in which a self-driving vehicle carrying multiple unmanned robots in the form of the mothership travels from a depot to a number of stations distributed in a neighborhood to perform multiple pickup and delivery services. First of all, we present it as a Multi-modal Vehicle Routing Problem (MMVRP) with time constraints, which are typical service requirements for grocery and food delivery in practice. We then formulate it as a Mixed Integer Quadratically Constrained Program (MIQCP) model to determine the optimal integrated routing plan (vehicle routing and robot routing) to minimize the total weighted tardiness of all services. Finally, a small-size and a medium-size problem instance are solved using the Gurobi solver in Python to demonstrate the validity and the performance of the proposed MIQCP model.

*Keywords*-vehicle routing; pickup and delivery; time constraints; mothership system; mixed integer quadratic programming.

## I. INTRODUCTION

Last-mile delivery is undergoing dramatic and rapid changes with the advance of the technologies such as self-driving cars and autonomous robots [1]. DoorDash, the largest food delivery company in the United States, has been using unmanned delivery robots to fulfill food orders since 2017. DoorDash's robots can carry 22 pounds of food. During the delivery, DoorDash's robots keep the food cabin locked to ensure its safety. When the robot arrives at the destination, DoorDash sends a text message with a link to its customer. The customer clicks the link to unlock the food cabin. In addition to flat roads, robots have recently been used to deliver meals across floors of office buildings in Shanghai, China. These robots can take the elevator all by themselves without manual intervention with built-in autonomous positioning and navigation system. It is reported [2] that these delivery robots significantly alleviate the peak lunch crowd and save one-third of the delivery time.

China's express delivery sector grows with e-commerce and is the largest express delivery market by volume in the world. According to the State Post Bureau [3], this sector is expected to handle 95 billion parcels and generate RMB 1 trillion (about 154.6 billion U.S. dollars) in revenue in 2021. However, the rapidly increasing labor costs and recent government policies to protect couriers' rights make Chinese delivery firms no easy victories. Lately, Chinese delivery firms have shown tremendous desire to use new technologies to make their business models more sustainable. Because of the close tie between e-commerce and express delivery, almost all Chinese e-commerce giants, such as Meituan (backed by Tencent), Alibaba, and JD.com, have initiated their unmanned delivery solutions in 2021. For example, in April of 2021, Meituan launched its new Modai 20 unmanned delivery vehicle in Beijing, Shanghai, and Shenzhen. Modai 20 can deliver parcels or meals within closed or semi-controlled environments, such as office complexes and school campuses. It is clear that the need for intelligent, unmanned last-mile delivery solutions is growing.

Within this context, this paper first conceptualizes an unmanned vehicle-robot pickup and delivery system, in which a self-driving vehicle carrying multiple unmanned robots in the form of the mothership travels from a depot to a number of stations to perform multiple pickup and/or delivery services. We then present it as a Multi-modal Vehicle Routing Problem (MMVRP) with time constraints, which are typical service requirements for grocery and food delivery in practice. After that, we formulate it as a Mixed Integer Quadratically Constrained Program (MIQCP) model to determine the optimal integrated routing plan (vehicle routing and robot routing) to minimize the total weighted tardiness of all services with heterogeneous importance. To the best of the authors' knowledge, this paper is the first study considering time constraints in the routing integration of multi-modal pickup and delivery problems in the existing literature.

The remainder of this paper is organized as follows. The most relevant literature in vehicle routing is reviewed in Section 2. After that, the problem description and mathematical formulation are presented in detail in Section 3. In Section 4, a small-size and a medium-size problem instance are solved to demonstrate the validity and the performance of the proposed optimization model. The conclusions and future work are summarized in Section 5.

## II. Literature Review

Vehicle Routing Problem (VRP) is one of the most classical and intensively studied optimization problems because of its many practical applications in the real world. Since its first introduction in the paper of George Dantzig and John Ramser in 1959 [4], many VRP variants have been proposed and studied in the extant literature [5], [6]. Among which, the most relevant VRP variants to our study are Vehicle Routing Problem with Pickup and Delivery (VRPPD) [7], Vehicle Routing Problem with Time Windows (VRPTW) [8], and Multi-modal Vehicle Routing Problem (MMVRP).

Vehicle Routing Problem with Pickup and Delivery (VRPPD) considers the situation where goods need to be picked up from a specific location and dropped off at their designated destination [5]. In the VRPPD, the pickup and drop-off must be done by the same vehicle typically; therefore the pickup location and drop-off location must be included in the same route [7], [9].

Vehicle Routing Problem with Time Windows (VRPTW) considers the situation where deliveries made to each customer must occur in a specified time interval [5]. This time constraint can be treated as hard as a strict constraint [10], [11] or soft with a penalty cost [12], [13].

Multi-modal Vehicle Routing Problem (MMVRP) refers to the type of VRP, which involves at least two transportation modes (e.g., truck and robot) and therefore requires coordination between different transportation resources. Lin [14] studied an MMVRP with a heavy resource (truck) and a light resource (foot couriers). The author developed a mixed integer programming (MIP) formulation to minimize the total cost.

Some studies consider vehicle-drone combined operations. Murray and Chu [15] introduced a sidekick truck-drone delivery system (one drone is mounted on a truck) to visit customers. Two MIP formulations and two heuristics are proposed to minimize the delivery completion time, which is further tested on problem instances up to 10 customers. Later, Agatz et al. [16] proposed a truck-first-drone-second heuristic to solve a similar Vehicle Routing Problem with Drone (VRP-D). Wang et al. [17] and Poikonen et al. [18] discussed the worst-case bounds for a more general form of VRP-D with multiple trucks and drones without an optimization framework. Ham [19] proposed a constraint programming method to solve the VRP-D to minimize the total operation cost. Kim and Moon [20] presented a truck-drone system with drone stations to overcome the flight-range limitation. For a similar sidekick system, Jeong et al. [21] considered drone energy consumption and restricted flying areas, and Gonzalez-R et al. [22] considered the limited power-life of batteries. Kitjacharoenchai, Min, and Lee [23] model multiple truck and drone systems as two-echelon vehicle routing problems, where the first level deals with truck routing and the second level deal with the drones routing. Recently, Karak and Abdelghany [24] introduced a full vehicle-drone routing integration in the form of a mothership system, where the "swarm" dispatching allows multiple pickups and deliveries to be performed simultaneously.

Some studies consider vehicle-robot combined operations. Boysen, Schwerdfeger, and Weidinger [25] considered a truck-robot delivery system where each robot is dedicated to a single customer. They developed a MIP formulation and derived heuristics to solve it. Simonia, Kutanoglub and Claudela [26] considered a truck-robot pickup and delivery system without time-windows, in which both the truck and robots can visit customers. The authors adopted iterated local search with an adaptive perturbation (LS-AP) to explore possible transportation routes. Recently, Chen et al. [27], [28] developed meta heuristic and search heurisitic for a vehicle-robot delivery problem with time windows, in which each robot is allowed to deliver once at most at each station.

So far, to the best of our knowledge, this paper is the first study considering routing integration for a Multi-modal pickup and delivery system with time constraints in the form of the mothership. We extend Karak and Abdelghany's study [24] by considering time constraints for pickup and delivery services, which are typical requirements for grocery and food delivery in practice.

## III. Mathematical Modeling

### A. Problem Description

This section presents the routing problem of the conceptualized vehicle-robot pickup and delivery system. The following assumptions are made to specify the operation characteristics of the proposed mothership system.

1) Multiple robots are mounted on a single vehicle.
2) Each station can be visited by the vehicle only once.
3) Customers are served only by robots.
4) Multiple robots can be dispatched simultaneously from any station to enhance the overall productivity of the system.
5) Packages in standard insulated containers are loaded and unloaded automatically from the robots once the robots have arrived at customers' locations or returned to the vehicle.
6) At each station, each robot can perform multiple delivery and/or pickup services as long as its travel range is not violated. However, each robot has to return to the dispatching station to load and unload containers between two consecutive services.
7) Each robot must be dispatched and collected at the same station.
8) Each service has its importance factor, indicating the importance of this service.
9) The vehicle cannot move from a station before all robots return to that station.

10) Robot batteries are replaced with fully charged batteries each time they are collected by the vehicle.

Next, a mathematical formulation in the form of a MIQCP is developed for the routing problem described above.

### B. Notation

**Sets:**

$\mathcal{S}$ = Set of stations =$1, 2, ..., n_s$, denoted by subscript $k, l, m$
$\mathcal{R}$ = Set of robots =$1, 2, ...n_r$, denoted by subscript $r$
$\mathcal{C}$ = Set of customers =$1, 2, ...n_c$, denoted by subscript $o, p, q$

**Parameters:**

$T_o$ = Deadline of the service of customer $o$
$WI_o$ = The importance of customer $o$
$TR$ = Maximum travel range of robot $r \in R$
$L_{kl}$ = The length of link $(k, l)$, $k \in S$ and $l \in S$, distances between stations
$L_{ko}$ = The length of link $(k, o)$, $k \in S$ and $o \in C$, distances from stations to customers
$L_{op}$ = The length of link $(o, p)$, $o \in C$ and $p \in C$, distances between customers
$VV$ = Travel speed of the vehicle
$VR$ = Travel speed of the robot

**Decision Variables:**

$y_{kl}$ = 1 if vehicle travels link $(k, l)$, and 0 otherwise
$x_{rko}$ = 1 if robot $r$ dispatched at station $k$ and travels on link $(k, o)$, and 0 otherwise
$z_{rok}$ = 1 if robot $r$ collected at station $k$ and travels on link $(o, k)$, and 0 otherwise
$w_{rkop}$ = 1 if robot $r$ dispatched at station $k$ and serves customer $p$ after customer $o$, and 0 otherwise
$t_k^{arrive}$ = Time arriving station $k$
$t_k^{depart}$ = Time leaving station $k$
$t_o^{complete}$ = Completion time of the service for customer $o$
$t_o^{tardiness}$ = Tardiness of the service for customer $o$

### C. Optimization Model

$$\min \sum_{o \in C} WI_o t_o^{tardiness}$$

subject to:

Depot constraints:

$$\sum_{k \in S} y_{0k} = 1 \quad (1)$$

$$\sum_{k \in S} y_{k0} = 1 \quad (2)$$

Station constraints:

$$\sum_{i \in 0 \cup S} y_{ik} = 1 \quad \forall k \in S \quad (3)$$

$$\sum_{i \in 0 \cup S} y_{ki} = 1 \quad \forall k \in S \quad (4)$$

$$\sum_{o \in C} x_{rko} \leq 1 \quad \forall k \in S, r \in R \quad (5)$$

$$\sum_{o \in C} z_{rok} \leq 1 \quad \forall k \in S, r \in R \quad (6)$$

$$\sum_{o \in C} x_{rko} = \sum_{o \in C} z_{rok} \quad \forall r \in R, k \in S \quad (7)$$

$$x_{rko} + \sum_{p \in C} w_{rkpo} = z_{rok} + \sum_{p \in C} w_{rkop} \quad \forall k \in S, r \in R, o \in C \quad (8)$$

Robot constraints:

$$\sum_{o \in C} x_{rko} \times L_{ko} + \sum_{o \in C} \sum_{p \in C} w_{rkop} \times (L_{ok} + L_{kp})$$
$$+ \sum_{o \in C} z_{rok} \times L_{ok} \leq TR \quad \forall r \in R, k \in S \quad (9)$$

Customer constraints:

$$\sum_{r \in R} \sum_{k \in S} x_{rkp} + \sum_{r \in R} \sum_{k \in S} \sum_{o \in C} w_{rkop} = 1 \quad \forall p \in C \quad (10)$$

$$\sum_{r \in R} \sum_{k \in S} z_{rok} + \sum_{r \in R} \sum_{k \in S} \sum_{p \in C} w_{rkop} = 1 \quad \forall o \in C \quad (11)$$

Time constraints:

$$t_0^{depart} = 0 \quad (12)$$

$$t_k^{arrive} = \sum_{i \in 0 \cup S} y_{ik} \times (t_i^{depart} + \frac{L_{ik}}{VV}) \quad \forall k \in S \quad (13)$$

$$t_k^{arrive} \leq t_k^{depart} \quad \forall k \in S \quad (14)$$

$$t_p^{complete} = \sum_{r \in R} \sum_{k \in S} x_{rkp} \times (t_k^{arrive} + \frac{L_{kp}}{VR})$$
$$+ \sum_{r \in R} \sum_{k \in S} \sum_{o \in C} w_{rkop} \times (t_o^{complete} + \frac{L_{ok} + L_{kp}}{VR}) \quad \forall p \in C \quad (15)$$

$$t_k^{depart} \geq z_{rok} \times (t_o^{complete} + \frac{L_{ok}}{VR}) \quad \forall k \in S, r \in R, o \in C \quad (16)$$

$$T_o \geq t_o^{complete} - t_o^{tardiness} \quad \forall o \in C \quad (17)$$

The objective is to minimize the total weighted tardiness of all pickup and delivery services.

Constraints (1) and (2) ensure that the vehicle departs from and returns to the depot to and from exactly one station. Constraints (3) and (4) enforce that the vehicle can only enter or exit any station once, whether it is from the depot or other stations. Constraints (5) and (6) make sure that all robots can only be dispatched or collected from any station at most once. Constraint (7) guarantees that the number of dispatch and collection of any robot at any station is equal. Constraint (8) ensures the continuity of the route constructed for each robot at each station. Constraint (9) ensures that the travel range of any robot will not be violated. Constraints (10) and (11) ensure that the incoming and outgoing flows of all robots to and from any consumer node should be exactly equal to 1. Constraint (12) initializes the start time of service of the whole mothership system, which is 0. Constraint (13) computes the arrival time of the vehicle to any station. Constraint (14) guarantees the arrival time of the vehicle at each station is earlier than its departure time. Constraint (15) calculates the service completion time of all customers. Constraint (16) enforces that the vehicle must wait for all robots to return before leaving any station. Constraint (17) measures the tardiness of each service.

## IV. NUMERICAL EXAMPLES

Two features characterize the complexity of the presented optimization problem. First, it is well known that the classic VRP is an NP-hard problem [29]. Second, the nonlinear constraints (15), (16), (17) make our problem an MIQCP, which can not be solved by using the CPLEX solver even for small-size problems. Due to such complexity, the size of problems that can be solved optimally is very limited. Next, a small-size and a medium-size problem instances are solved to demonstrate the validity and the performance of the proposed optimization model. The small-size problem instance can help explain intuitively the optimization model and the optimal solutions. The medium-size problem instance can help show the computational efficiency of the proposed optimization model. For these two problem instances, the optimization model described in Section III is implemented using the Python programming language and solved by the Gurobi solver.

### A. Small-size Problem

The small-size problem instance and its optimal routes (solutions) of the vehicle and robots are shown in Fig. 1. This small-size problem instance can represent the scenario of the delivery and pick up service for a small-size community or town. In this problem, we have one depot and two stations $\mathcal{S} = \{0, 1, 2\}$, two robots $\mathcal{R} = \{0, 1\}$, and eight customers $\mathcal{C} = \{0, 1, 2, 3, ...7\}$. The depot is depicted by the square filled with black color. The stations are depicted by the circles filled with gray color. The pentagons are to represent the customers. The optimal routes of the vehicle is represented by black solid lines. The optimal routes of the robots are represented by blue and red dashed lines. The different colors of the dashed lines are to distinguish the routes of different robots. Same conventions in Fig. 1 are also used to illustrate the problem and results in Fig. 2. The maximum travel range of the robot is $TR = 200$ miles. The travel speed of the vehicle is $VV = 50$ miles / hour, and the travel speed of the robot is $VR = 5$ miles / hour. The locations of the depot, stations, and customers are specified by their (x, y) coordinates in a 100 miles $\times$ 100 miles region. The depot is located at the coordinate (0, 0). The stations are evenly distributed in this region. And the (x, y) coordinates of all the customers are generated randomly using Python. In addition, the importance factors and the deadlines of each service are randomly generated between (0, 1) and [10, 50], respectively. All these parameter values are listed in Table I and Table II. For this small-size problem instance, there are in total 358 decision variables and 122 constraints in the optimization model.

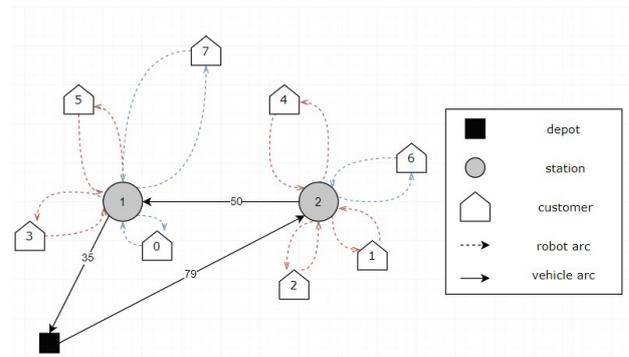

Figure 1. Small-size problem instance and the optimal routes.

Table I
LOCATIONS OF DEPOT AND STATIONS FOR SMALL-SIZE PROBLEM

| Depot / Station Name | Index | x | y |
|---|---|---|---|
| Depot | 0 | 0 | 0 |
| Station-1 | 1 | 25 | 25 |
| Station-2 | 2 | 75 | 25 |

Table II
PARAMETERS OF CUSTOMERS FOR SMALL-SIZE PROBLEM

| Customer Name | Index | x | y | Importance | Deadline [min] |
|---|---|---|---|---|---|
| Customer-0 | 0 | 19 | 39 | 0.36 | 18 |
| Customer-1 | 1 | 86 | 17 | 0.06 | 40 |
| Customer-2 | 2 | 87 | 1 | 0.55 | 21 |
| Customer-3 | 3 | 0 | 46 | 0.83 | 23 |
| Customer-4 | 4 | 70 | 58 | 0.87 | 10 |
| Customer-5 | 5 | 25 | 69 | 0.51 | 22 |
| Customer-6 | 6 | 80 | 90 | 0.95 | 16 |
| Customer-7 | 7 | 62 | 83 | 0.31 | 49 |

The optimal solutions attained are listed in Table III. For ease of illustration, we only list the non-zero solutions of the binary decision variables $y_{kl}, x_{rko}, z_{rok}, w_{rkop}$ in Table

Table III
OPTIMAL SOLUTION OF SMALL-SIZE PROBLEM

| Variable | Solution | Variable | Solution |
|---|---|---|---|
| $y_{0,2}$ | 1 | $t_1^{arrive}$ | 32.11 [min] |
| $y_{1,0}$ | 1 | $t_2^{arrive}$ | 1.58 [min] |
| $y_{2,1}$ | 1 | $t_1^{depart}$ | 65.72 [min] |
| $x_{0,1,3}$ | 1 | $t_2^{depart}$ | 31.11 [min] |
| $x_{0,2,4}$ | 1 | $t_3^{depart}$ | 35.15 [min] |
| $x_{1,1,0}$ | 1 | $t_0^{tardiness}$ | 17.15 [min] |
| $x_{1,2,6}$ | 1 | $t_1^{complete}$ | 28.39 [min] |
| $z_{0,1,2}$ | 1 | $t_2^{complete}$ | 20.30 [min] |
| $z_{0,5,1}$ | 1 | $t_3^{complete}$ | 38.64 [min] |
| $z_{1,6,2}$ | 1 | $t_3^{tardiness}$ | 15.64 [min] |
| $z_{1,7,1}$ | 1 | $t_4^{complete}$ | 8.26 [min] |
| $w_{0,1,3,5}$ | 1 | $t_5^{complete}$ | 53.97 [min] |
| $w_{0,2,2,1}$ | 1 | $t_5^{tardiness}$ | 31.97 [min] |
| $w_{0,2,4,2}$ | 1 | $t_6^{complete}$ | 14.62 [min] |
| $w_{1,1,0,7}$ | 1 | $t_7^{complete}$ | 51.96 [min] |
| Objective Solution | 36.52 [min] | $t_7^{tardiness}$ | 2.96 [min] |
| Solver Time | 11.09 [s] | | |

Table IV
OPTIMAL ROBOTS' ROUTES OF SMALL-SIZE PROBLEM

| Station | Robot Index | Route |
|---|---|---|
| S-1 | 0 | {S-1, C-3, S-1, C-5, S-1} |
| S-2 | 1 | {S-1, C-0, S-1, C-7, S-1} |
| S-1 | 0 | {S-2, C-4, S-2, C-2, S-2, C-1, S-2} |
| S-2 | 1 | {S-2, C-6, S-2} |

III. In other words, the solutions of the non-listed binary variables are equal to zero. We also list the solutions of the departure time $t_k^{depart}$, arrival time $t_k^{arrive}$, tardiness $t_o^{tardiness}$ and completion time $t_o^{complete}$ for all the customers. For this small-size problem instance, the minimal total weighted tardiness is 35.52. One personal computer running on the macOS Big Sur operating system with Intel Core-i5 1.6 GHz CPU and 4 GB RAM is deployed to solve the optimization model. It takes the Gurobi solver 11.09 seconds to find the optimal solution. The solved optimal vehicle's route is {Depot-0, Station-2, Station-1, Depot-0}. The solved optimal robots' routes are listed in Table IV, in which 'S-1' denotes Station-1, and 'C-3' denotes Customer-3. As it can be seen, all the robots return to the corresponding dispatching stations after finishing their pickup and delivery services.

### B. Medium-size Problem

The medium-size problem instance and its optimal solution are shown in Fig 2, in which there are 4 stations, 4 robots, and 12 customers. This medium-size problem instance can represent the scenario of the delivery and pick up service for a medium-size community or town. Since there are more robots in this problem instance, aside from blue and red, more colours (green and black) of the dashed lines are used to distinguish the routes of different robots. The maximum travel range of the robots is set as 80 miles.

Details of the parameter values of stations and customers are listed in Table V and Table VI, respectively. In this medium-size problem totally, we have 503 constraints and 816 decision variables in the optimization model.

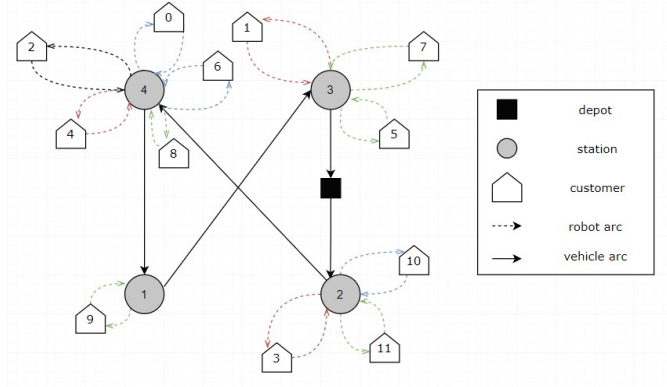

Figure 2. Medium-size problem instance and the optimal routes.

Table V
LOCATIONS OF DEPOT AND STATIONS FOR MEDIUM-SIZE PROBLEM

| Depot / Station Name | Index | x | y |
|---|---|---|---|
| Depot | 0 | 0 | 0 |
| Station-1 | 1 | 25 | 25 |
| Station-2 | 2 | 75 | 25 |
| Station-3 | 3 | 75 | 75 |
| Station-4 | 4 | 25 | 75 |

Table VI
PARAMETERS OF CUSTOMERS FOR MEDIUM-SIZE PROBLEM

| Customer Name | Index | x | y | Importance | Deadline [min] |
|---|---|---|---|---|---|
| Customer-0 | 0 | 48 | 71 | 0.55 | 37 |
| Customer-1 | 1 | 75 | 61 | 0.20 | 45 |
| Customer-2 | 2 | 1 | 98 | 0.63 | 19 |
| Customer-3 | 3 | 60 | 35 | 0.47 | 13 |
| Customer-4 | 4 | 17 | 58 | 0.01 | 36 |
| Customer-5 | 5 | 74 | 58 | 0.53 | 38 |
| Customer-6 | 6 | 80 | 90 | 0.96 | 39 |
| Customer-7 | 7 | 99 | 79 | 0.50 | 37 |
| Customer-8 | 8 | 43 | 45 | 0.78 | 47 |
| Customer-9 | 9 | 39 | 37 | 0.34 | 34 |
| Customer-10 | 10 | 55 | 23 | 0.87 | 47 |
| Customer-11 | 11 | 56 | 31 | 0.42 | 11 |

For this medium-size problem instance, the minimum total weighted tardiness is 1.47. One personal computer running on the macOS Big Sur operating system with Intel Core-i5 1.6 GHz CPU and 4 GB RAM is deployed to solve the optimization model. It takes Gurobi 342.49 seconds to find the optimal solution. As it is shown in Table VII, the optimal vehicle routing is {Depot-0. Station-2, Station-4, Station-1, Station-3, Station-0}. And the optimal robot routing for each robot at each station is presented in Table VIII.

Table VII
OPTIMAL SOLUTION OF MEDIUM-SIZE PROBLEM

| Variable | Solution | Variable | Solution |
| --- | --- | --- | --- |
| $y_{0,2}$ | 1 | $z_{3,9,1}$ | 1 |
| $y_{2,4}$ | 1 | $z_{3,11,2}$ | 1 |
| $y_{4,1}$ | 1 | $w_{0,4,6,0}$ | 1 |
| $y_{1,3}$ | 1 | $t_1^{arrive}$ | 26.03 [min] |
| $y_{3,0}$ | 1 | $t_2^{arrive}$ | 1.58 [min] |
| $x_{0,2,10}$ | 1 | $t_3^{arrive}$ | 34.82 [min] |
| $x_{0,4,6}$ | 1 | $t_4^{arrive}$ | 11.04 [min] |
| $x_{1,2,3}$ | 1 | $t_1^{depart}$ | 33.41 [min] |
| $x_{1,3,1}$ | 1 | $t_2^{depart}$ | 9.62 [min] |
| $x_{1,4,4}$ | 1 | $t_3^{depart}$ | 44.55 [min] |
| $x_{2,3,7}$ | 1 | $t_4^{depart}$ | 25.03 [min] |
| $x_{2,4,2}$ | 1 | $t_0^{complete}$ | 19.70 [min] |
| $x_{1,3,9}$ | 1 | $t_1^{complete}$ | 37.62 [min] |
| $x_{3,2,11}$ | 1 | $t_2^{complete}$ | 17.68 [min] |
| $x_{3,3,5}$ | 1 | $t_3^{complete}$ | 5.19 [min] |
| $x_{3,4,8}$ | 1 | $t_4^{complete}$ | 17.63 [min] |
| $z_{0,0,4}$ | 1 | $t_5^{complete}$ | 38.23 [min] |
| $z_{0,10,12}$ | 1 | $t_6^{complete}$ | 13.04 [min] |
| $z_{1,1,3}$ | 1 | $t_7^{complete}$ | 39.69 [min] |
| $z_{1,3,2}$ | 1 | $t_8^{complete}$ | 18.03 [min] |
| $z_{1,4,4}$ | 1 | $t_9^{complete}$ | 29.72 [min] |
| $z_{2,2,4}$ | 1 | $t_{10}^{complete}$ | 5.60 [min] |
| $z_{2,7,3}$ | 1 | $t_{11}^{complete}$ | 5.57 [min] |
| $z_{3,5,3}$ | 1 | $t_5^{tardiness}$ | 0.23 [min] |
| $z_{3,8,4}$ | 1 | $t_7^{tardiness}$ | 2.69 [min] |
| Objective Solution | 1.47 | Solver Time | 342.49 [s] |

Table VIII
OPTIMAL ROBOTS' ROUTES OF MEDIUM-SIZE PROBLEM

| Station | Robot Index | Route |
| --- | --- | --- |
| S-2 | 0 | {S-2, C-10, S-2} |
| S-2 | 1 | {S-2, C-3, S-2} |
| S-2 | 3 | {S-2, C-11, S-2} |
| S-4 | 0 | {S-4, C-6, S-4, C-0, S-4} |
| S-4 | 1 | {S-4, C-4, S-4} |
| S-4 | 2 | {S-4, C-2, S-4} |
| S-4 | 3 | {S-4, C-8, S-4} |
| S-1 | 3 | {S-1, C-9, S-1} |
| S-3 | 1 | {S-3, C-1, S-3} |
| S-3 | 3 | {S-3, C-7, S-3} |
| S-3 | 3 | {S-3, C-5, S-3} |

## V. CONCLUSIONS AND FUTURE WORK

This paper considers the problem of dispatching a mothership system including one vehicle mounted with multiple robots from depot and stations for the pickup and delivery services to multiple customers. The vehicle travels in between the depot and stations. The robots delivers and picks up the goods to the customers. The depot, stations and customers are spread in different locations. This problem is to represent the actual last-mile delivery service using self-driving cars and autonomous robots in the e-commerce sector. The objective is to minimize the total weighted tardiness of all customers. The considered constraints include the time windows of pickup and delivery services, the travel ranges of the robots, and the practical requirements and characteristics of the mothership system. The vehicle-robot routing problem is formulated as a MIQCP model and implemented in the Python programming language. Two problem instances are optimally solved by the Gurobi solver. The optimal routing of the vehicle and the robots are found within acceptable computing time. Future research work dedicated to more problem instances or more efficient solution methodologies, such as search heuristics and metaheuristics, is expected.

REFERENCES

[1] F. Wang, F. Wang, X. Ma, and J. Liu, "Demystifying the crowd intelligence in last mile parcel delivery for smart cities," *IEEE Network*, vol. 33, no. 2, pp. 23–29, 2019.

[2] Sohu. (2020) Shanghai office robots take the elevator to deliver meals by themselves. Https://www.sohu.com/a/419255840114778.

[3] W. Luo. (2021) Express delivery sector sees rapid development. Https://www.chinadaily.com.cn/a/202103/27.

[4] G. B. Dantzig and J. H. Ramser, "The truck dispatching problem," *Management science*, vol. 6, no. 1, pp. 80–91, 1959.

[5] K. Braekers, K. Ramaekers, and I. Van Nieuwenhuyse, "The vehicle routing problem: State of the art classification and review," *Computers & Industrial Engineering*, vol. 99, pp. 300–313, 2016.

[6] B. Eksioglu, A. V. Vural, and A. Reisman, "The vehicle routing problem: A taxonomic review," *Computers & Industrial Engineering*, vol. 57, no. 4, pp. 1472–1483, 2009.

[7] A. S. Tasan and M. Gen, "A genetic algorithm based approach to vehicle routing problem with simultaneous pick-up and deliveries," *Computers & Industrial Engineering*, vol. 62, no. 3, pp. 755–761, 2012.

[8] M. M. Solomon, "Vehicle routing and scheduling with time window constraints: Models and algorithms," Tech. Rep., 1984.

[9] L. Pradenas, B. Oportus, and V. Parada, "Mitigation of greenhouse gas emissions in vehicle routing problems with backhauling," *Expert Systems with Applications*, vol. 40, no. 8, pp. 2985–2991, 2013.

[10] A. Agra, M. Christiansen, R. Figueiredo, L. M. Hvattum, M. Poss, and C. Requejo, "The robust vehicle routing problem with time windows," *Computers & operations research*, vol. 40, no. 3, pp. 856–866, 2013.

[11] T. Vidal, T. G. Crainic, M. Gendreau, and C. Prins, "A hybrid genetic algorithm with adaptive diversity management for a large class of vehicle routing problems with time-windows," *Computers & operations research*, vol. 40, no. 1, pp. 475–489, 2013.

[12] M. A. Figliozzi, "An iterative route construction and improvement algorithm for the vehicle routing problem with soft time windows," *Transportation Research Part C: Emerging Technologies*, vol. 18, no. 5, pp. 668–679, 2010.

[13] D. Taş, N. Dellaert, T. Van Woensel, and T. De Kok, "Vehicle routing problem with stochastic travel times including soft time windows and service costs," *Computers & Operations Research*, vol. 40, no. 1, pp. 214–224, 2013.

[14] C. Lin, "A vehicle routing problem with pickup and delivery time windows, and coordination of transportable resources," *Computers & Operations Research*, vol. 38, no. 11, pp. 1596–1609, 2011.

[15] C. C. Murray and A. G. Chu, "The flying sidekick traveling salesman problem: Optimization of drone-assisted parcel delivery," *Transportation Research Part C: Emerging Technologies*, vol. 54, pp. 86–109, 2015.

[16] N. Agatz, P. Bouman, and M. Schmidt, "Optimization approaches for the traveling salesman problem with drone," *Transportation Science*, vol. 52, no. 4, pp. 965–981, 2018.

[17] X. Wang, S. Poikonen, and B. Golden, "The vehicle routing problem with drones: several worst-case results," *Optimization Letters*, vol. 11, no. 4, pp. 679–697, 2017.

[18] S. Poikonen, X. Wang, and B. Golden, "The vehicle routing problem with drones: Extended models and connections," *Networks*, vol. 70, no. 1, pp. 34–43, 2017.

[19] A. M. Ham, "Integrated scheduling of m-truck, m-drone, and m-depot constrained by time-window, drop-pickup, and m-visit using constraint programming," *Transportation Research Part C: Emerging Technologies*, vol. 91, pp. 1–14, 2018.

[20] S. Kim and I. Moon, "Traveling salesman problem with a drone station," *IEEE Transactions on Systems, Man, and Cybernetics: Systems*, vol. 49, no. 1, pp. 42–52, 2018.

[21] H. Y. Jeong, B. D. Song, and S. Lee, "Truck-drone hybrid delivery routing: Payload-energy dependency and no-fly zones," *International Journal of Production Economics*, vol. 214, pp. 220–233, 2019.

[22] P. L. Gonzalez-R, D. Canca, J. L. Andrade-Pineda, M. Calle, and J. M. Leon-Blanco, "Truck-drone team logistics: A heuristic approach to multi-drop route planning," *Transportation Research Part C: Emerging Technologies*, vol. 114, pp. 657–680, 2020.

[23] P. Kitjacharoenchai, B.-C. Min, and S. Lee, "Two echelon vehicle routing problem with drones in last mile delivery," *International Journal of Production Economics*, vol. 225, p. 107598, 2020.

[24] A. Karak and K. Abdelghany, "The hybrid vehicle-drone routing problem for pick-up and delivery services," *Transportation Research Part C: Emerging Technologies*, vol. 102, pp. 427–449, 2019.

[25] N. Boysen, S. Schwerdfeger, and F. Weidinger, "Scheduling last-mile deliveries with truck-based autonomous robots," *European Journal of Operational Research*, vol. 271, no. 3, pp. 1085–1099, 2018.

[26] M. D. Simoni, E. Kutanoglu, and C. G. Claudel, "Optimization and analysis of a robot-assisted last mile delivery system," *Transportation Research Part E: Logistics and Transportation Review*, vol. 142, p. 102049, 2020.

[27] C. Chen, E. Demir, and Y. Huang, "An adaptive large neighborhood search heuristic for the vehicle routing problem with time windows and delivery robots," *European Journal of Operational Research*, 2021.

[28] C. Chen, E. Demir, Y. Huang, and R. Qiu, "The adoption of self-driving delivery robots in last mile logistics," *Transportation research part E: logistics and transportation review*, vol. 146, p. 102214, 2021.

[29] B. L. Golden, S. Raghavan, and E. A. Wasil, *The vehicle routing problem: latest advances and new challenges*. Springer Science & Business Media, 2008, vol. 43.